\documentclass[submit]{smj}

\usepackage{amsmath}
\usepackage{amsthm}
\usepackage{amssymb}
\usepackage{color}
\usepackage{graphicx,psfrag,epsf}
\usepackage{enumerate}
\usepackage{natbib}
\usepackage{subfigure}
\usepackage{url} 

\newcommand{\M}[1]{\boldsymbol{#1}}  
\newcommand{\V}[1]{\boldsymbol{#1}}  
\newcommand{\Unif}[0]{\textrm{Uniform}}

\renewcommand{\mod}[0]{\textrm{ mod }}

\Author{Ali Esmaieeli Sikaroudi \Affil{1}, 
        and Chiwoo Park \Affil{2}
}
\AuthorRunning{Esmaieeli and Park}

\Affiliations{

\item JP Morgan Chase and Co., 
      Jacksonville,
      FL, USA

\item Department of Industrial and Manufacturing Engineering,
      Florida State University, 
      Tallahassee,
      FL, USA
}   

\CorrAddress{Chiwoo Park, 
             Department of Industrial and Manufacturing Engineering,
                   Florida State University,
                   2525 Pottsdamer St., 
                   Tallahassee,
                   FL, USA}
\CorrEmail{cpark5@fsu.edu}
\CorrPhone{(+1)\;850\;410\;6457}
\CorrFax{(+1)\;850\;410\;6546}

\Title{A Mixture of Linear-Linear Regression Models for Linear-Circular Regression}
\TitleRunning{A Mixture of Linear-Linear Regression for Linear-Circular Regression}

\Abstract{
We introduce a new approach to a linear-circular regression problem that relates multiple linear predictors to a circular response. We follow a modeling approach of a wrapped normal distribution that describes angular variables and angular distributions and advances it for a linear-circular regression analysis.  Some previous works model a circular variable as projection of a bivariate Gaussian random vector on the unit square, and the statistical inference of the resulting model involves complicated sampling steps. The proposed model treats circular responses as the result of the modulo operation on unobserved linear responses. The resulting model is a mixture of multiple linear-linear regression models. We present two EM algorithms for maximum likelihood estimation of the mixture model, one for a parametric model and another for a non-parametric model. The estimation algorithms provide a great trade-off between computation and estimation accuracy, which was numerically shown using five numerical examples. The proposed approach was applied to a problem of estimating wind directions that typically exhibit complex patterns with large variation and circularity. 
}


\begin{document}

\maketitle

\section{INTRODUCTION}

Regression modeling with circular or directional data has application in numerous disciplines such as environmental fields including wind direction \citep{jammalamadaka2006effect} and wave direction  \citep{jona2012spatial}, polymer science  \citep{hamaide1991circular}, biology and medicine \citep{bell2008analysing,gao2006application,mooney2003fitting},  music \citep{dufour2015improving} and social science \citep{brunsdon2006using,gill2010circular}. Such applications have motivated scientists to develop regression models which are able to handle circular data. Depending on whether the predictors or responses are circular, the regression model are classified into a linear-circular model, a circular-linear model, and a circular-circular model, where `linear' implies variables defined on the real line topology and `circular' implies variables defined on the unit circle topology. This paper has focus on a linear-circular model, which relate linear predictors $X$ to a circular response $\Theta$. Although $X$ can be multivariate, we do not use a bold face font for $X$, because it can be confused with a matrix symbol.  

The main challenge in a linear-circular regression is the circular topology of the response variable, which makes the regression models proposed for a real response are not directly applicable for the linear-circular problem. Note that the distance between two real values can be quantified by the Euclidean distance in a real line, while it does not apply for two circular values due to the periodic nature of a circle, i.e., $\{2z\pi + \Theta: z \mbox{ is an integer}\}$ is the equivalent class of $\Theta$. The circularity issue was addressed  by three different approaches, the von Mises distribution \citep{gould1969regression, johnson1978some, fisher1992regression}, non-parametric circular regression \citep{di2013non}, and the projected linear model \citep{presnell1998projected,nunez2005bayesian,nunez2011bayesian, wang2013directional}. The wrapped normal model has been popular to describe a probability distribution of circular data \citep{jammalamadaka2001topics, mardia2009directional}, but its use for a regression analysis has been limited \citep{fisher1994time}. We briefly summarize these approaches below.

The very early researches on the linear-circular regression problem were mostly based on the assumption that the circular response $\Theta$ given predictors $X=\V{x}$ follows the von Mises distribution with the density,
\begin{equation}\label{Schroedinger}
f\left(\Theta;\mu(\V{x}),\kappa\right) = \dfrac{1}{2\pi I_{0}\left(\kappa\right)}\exp\{\kappa \cos\left(\Theta- \mu(\V{x})\right)\},
\end{equation}
where $I_{0}(\kappa)$ is the modified Bessel function of order 0, and $\mu(\V{x})$ and $\kappa$ are the mean and concentration parameters of the distribution. The circular mean $\mu(\V{x})$ is regressed over observations $\{(\V{x}_i, \theta_i)\}$ via circular link function $g$,
\begin{equation*}
\mu(\V{x}_i) = \mu_0 + g(\V{x}_i).
\end{equation*}
\citet{gould1969regression} proposed $g(\V{x}) = \V{\beta}' \V{x}$, and \citet{johnson1978some} used $g(\V{x})= 2 \pi F\left(\V{x}\right)$, where $F$ is the marginal distribution function of $\V{x}$. Later, \citet{fisher1992regression} proposed another form $g(\V{x}) = 2 \tan^{-1}\left(sgn\left(\V{\beta}' \V{x}\right)\vert \V{\beta}' \V{x} \vert^{\lambda}\right)$. Some maximum likelihood approaches were proposed to estimate the regression parameters. However, the likelihood functions of the proposed models are very difficult to optimize due to multi-modality having very narrow and sharp modes and unidentifiability of parameters \citep{presnell1998projected}.  

Another approach is to use a non-parametric smoothing  \citep{di2013non}. The approach is to find an unknown regression function $\mu(\cdot)$ that minimizes the angular risk function,
\begin{equation*}
E[1-\cos(\Theta-\mu(X))|X=\V{x}].
\end{equation*}
The minimizer of the risk is $\arctan(s(\V{x}), c(\V{x}))$, where $s(\V{x}) = E[\sin(\Theta)|X=\V{x}]$ and $c(\V{x}) = E[\cos(\Theta)|X=\V{x}]$. The non-parametric estimates of $s(\cdot)$ and $c(\cdot)$ were achieved using the locally weighted regression over $\{\sin(\theta_i)\}$ and $\{\cos(\theta_i)\}$ respectively, and the estimates were plugged into $\arctan(s(\V{x}), c(\V{x}))$ to give the estimate of $\mu(\V{x})$.

A more popular approach is to treat a circular response as the projection of unobserved bivariate normal variables on the unit circle \citep{presnell1998projected}, 
\begin{equation*}
\theta_i = \mbox{atan2}(y_{i2} / y_{i1}),
\end{equation*}
where $(y_{i1}, y_{i2})$ is a bivariate normal random vector with covariance $\M{\Sigma}$ and mean $\V{\mu}_i = \M{V}'\V{x}_i$. The conditional distribution of $\theta_i$ given $\V{x}_i$ is called the offset-normal distribution \citep{mardia2014statistics} or the projected normal distribution \citep{wang2013directional}. \citet{presnell1998projected} fixed $\M{\Sigma} = \M{I}$ and used the Expectation Maximization (EM) algorithm to estimate $\M{V}$. \citet{nunez2005bayesian} solved a version of the model using Bayesian approaches. More recently \citet{wang2013directional} analyzed a generalized version of the model with asymmetry and bimodality of the projected normal distribution. The model parameter estimation  requires computationally expensive Metropolis-Hastings samplings. \citet{nunez2011bayesian} proposed the Gibbs sampler for reduced computation. However, it still needs some Metropolis-Hastings sampling steps within the Gibbs sampler, and the number of the Metropolis-Hastings steps increases linearly in the number of observations, which makes the parameter estimation extremely slow for a large data size. \citet{hernandez2017general} introduced a new parameterization of the project normal distribution and proposed a slice sampler for a faster sampling.

Using a wrapped distribution is another popular approach. In the approach, a circular response $\Theta$ is regarded as the result of the modulo operation on a real random variable. When the real variable follows a probability distribution with density $f$, the corresponding circular response follows a wrapped distribution with the following density \citep[Section 3.5.7]{mardia2009directional},
\begin{equation} \label{eq:wrapped_model}
f_w(\Theta) = \sum_{Z=-\infty}^{\infty} f(\Theta + 2Z\pi + \pi).
\end{equation} 
The density function is achieved by wrapping the density $f$ around the circumference of an unit circle, and the name `wrapped' originated from the way the density function is obtained. The wrapped distribution has been a popular model of describing a probability distribution for a circular random variable. The $R$ package \texttt{Wrapped} is also available as open source software for the parameter estimation of the wrapped distribution. Many of the existing works in the wrapped normal distribution are concentrated on estimating the parameters of a wrapped distribution \citep{nodehi2018estimation}, but there are not many works related to a regression analysis for a circular response variable. The time series analysis of circular data with the wrapped normal and ARMA model was discussed \citep{fisher1994time}, and the EM algorithm was proposed to estimate the ARMA parameters. The parameter estimation involves large infinite sums, which makes the method computationally inefficient. 

In this paper, we follow a modeling approach of a wrapped normal distribution that describes an angular distribution and advances it for a linear-circular regression analysis. The new approach models a linear-circular regression with a mixture of linear-linear regression models, for which the identifiability and the statistical inference algorithm were well established. It is referred to \textit{the Angular Gaussian Mixture Model}, shortly the AGMM. The complexity of the model estimation is as simple as solving a standard Gaussian mixture model, so it is computationally feasible. The AGMM also provided more accurate estimation for many numerical examples. We will introduce the new model in Section 2. The statistical inference and the choice of tuning parameters will be discussed in Section 3. Five numerical examples will be presented with comparison to the projected linear model \citep{nunez2011bayesian} and the non-parametric smoothing \citep{di2013non} in Section 4. The application of the AGMM for a problem of estimating wind directions in time will be presented in Section \ref{sec:app}. Section 6 concludes this paper.

\section{AGMM Model}
Consider a general linear-circular regression problem for $p$ real predictors $X$ and a circular response $\Theta$. Following the wrapped normal model \citep{mardia2009directional}, we treat a circular response $\Theta \in [-\pi, \pi]$ as the result of the modulo operation on a  latent real (or linear) response $Y \in \mathbb{R}$,
\begin{equation*}
\Theta = (Y \mod 2\pi) - \pi. 
\end{equation*}
Equivalently, we can write
\begin{equation} \label{eq:lat_model}
Y = \Theta + 2Z\pi + \pi
\end{equation}
for an arbitrary integer $Z$. Consider the range of $Z$, $Z \in \{-K,\ldots, K\}$. For the time being, we assume $K$ is known. 

For a linear-circular regression, we impose a normal distribution on the unobserved response variable $Y$ given predictor variables $X=\V{x}$,
\begin{equation} \label{eq:norm}
Y|X=\V{x} \sim \mathcal{N}(\mu(\V{x}), \sigma^2(\V{x})),
\end{equation}
where $\mu(\cdot)$ and $\sigma^2(\cdot)$ are continuous functions. Let $f(y|\mu, \sigma^2)$ denote its density function. We further impose a discrete distribution over $Z|X=\V{x}$, 
\begin{equation}
P(Z=k|X=\V{x}) = r_k(\V{x}) \mbox{ for } k = -K,...,K,
\end{equation}
where $r_k(\V{x}) \in [0, 1]$ and $\sum_{k=-K}^K r_k(\V{x}) = 1$, and the conditional distribution of $\Theta$ conditioned on $Z=k$ and $X=\V x$ has the density,
\begin{equation*}
g(\theta|k, \V x) = \frac{f(\theta + 2k\pi + \pi|\mu, \sigma^2)}{r_k(\V{x})}.
\end{equation*}
Given those, the marginal distribution of $\Theta$ given $X=\V{x}$ has the following density,
\begin{equation*}
\begin{split}
h(\theta|\V x) & = \sum_{k=-K}^{K}  r_k(\V{x}) g(\theta|k, \V x) \\
			   &= \sum_{k=-K}^{K} f(\theta + 2k\pi + \pi|\mu, \sigma^2)
\end{split}
\end{equation*}
This can be seen as a truncated series of the wrapped normal model \eqref{eq:wrapped_model} being applied to a regression problem. Please note that $f$ is a normal density, and the last line of the density function expression can be also written as
\begin{equation*}
\begin{split}
h(\theta|\V x) = \sum_{k=-K}^{K} f(\theta|\mu_k, \sigma^2) 
\end{split}
\end{equation*}
where $\mu_k(\V x) = \mu(\V x) - 2k\pi - \pi$.  

\begin{figure}[t]
	\includegraphics[width = \linewidth]{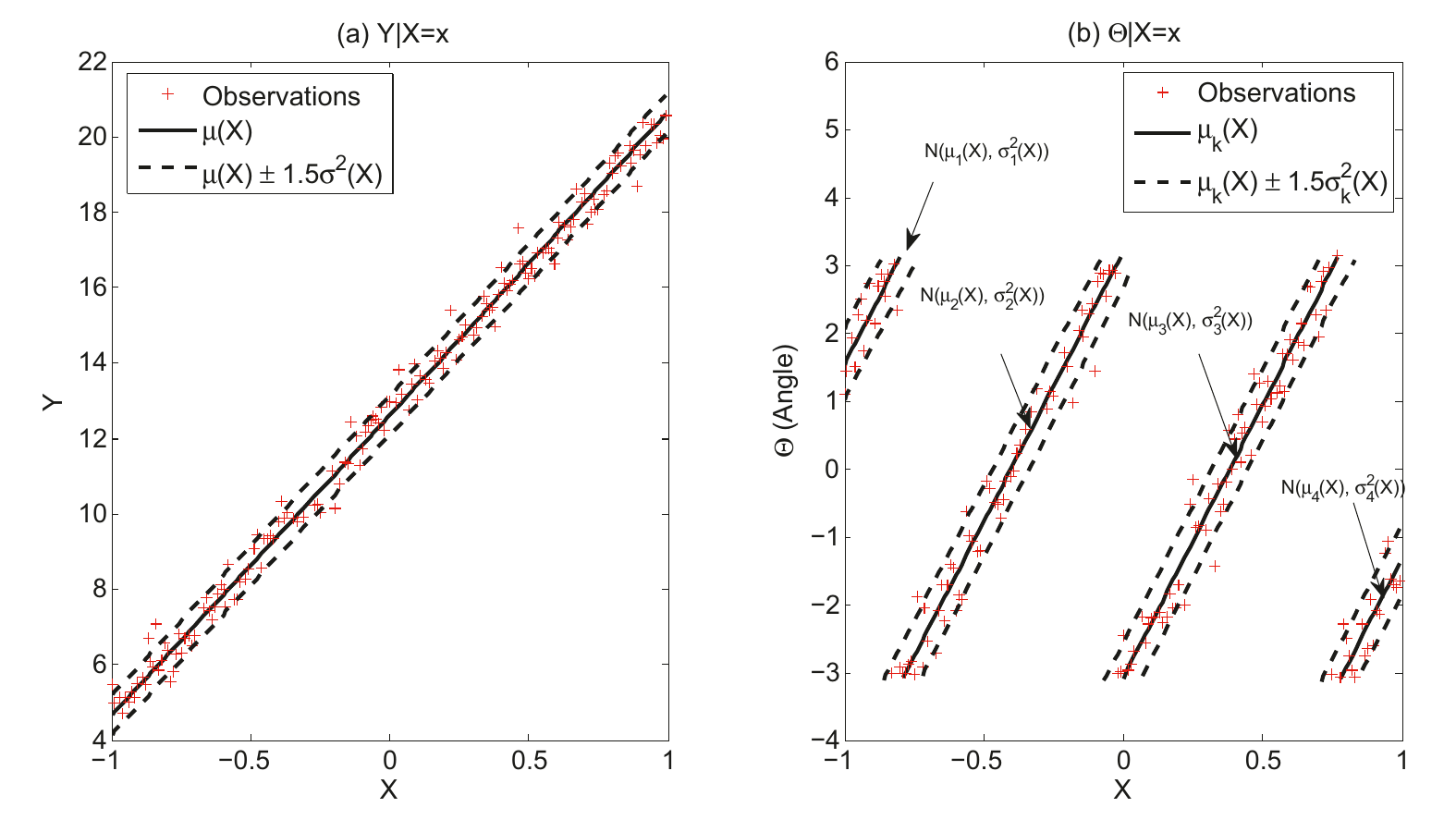}
	\centering
	\caption{Circular response can be seen as the result of the modulo operation on a linear response.}
	\label{fig:Figure1}
\end{figure}

One can assume certain parametric forms for $\mu(\V{x})$ and $\sigma^2(\V{x})$. For example, $\mu(\V{x}) = \V{B}(\V{x})' \V{\beta}$ and $\sigma^2(\V{x}) = \sigma^2$, where $\V{B}(\V{x})$ be the $q$-dimensional vector of nonlinear basis function values $B_j(\V{x})$'s, and $\V{\beta}$ is an $q$ dimensional vector of unknown coefficients. One can also consider $\mu(\V{x})$ and $\sigma^2(\V{x})$ as non-parametric functions. For the case, the model becomes a finite mixture of non-parametric regression models. In any case, the identifiability of the mixture model is well studied \citep[Theorem 1]{huang2013nonparametric}. 
\\
\noindent \textit{\textbf{Remark.} The AGMM is a mixture of linear-linear regression models. Figure \ref{fig:Figure1} illustrates an perspective to interpret AGMM. Note that the unobserved variable $Y|X=\V{x}$ comes from a multiple linear regression model $\mathcal{N}(\mu(\V{x}), \sigma^2(\V{x}))$, which typically forms a continuous regression line (Figure \ref{fig:Figure1}-(a)). However, due to the result of the modulo operation on $Y|X=\V{x}$, $\Theta|X=\V{x}$ exhibits discontinuity, so the random sample from $\Theta|X=\V{x}$ appears mixed observations from multiple different models $\mathcal{N}(\mu_k(\V{x}), \sigma^2(\V{x}))$ (Figure \ref{fig:Figure1}-(b)). The AGMM model fits a mixture of regression models to the observation to infer the unobserved model $Y|X=\V{x}$, so the $\Theta|X=\V{x}$ can be inferred by taking the modulo operator on $Y|X=\V{x}$. }

\section{Model Estimation}
Assume that $\{(\V{x}_i, \theta_i), i = 1,...,n\}$ is a random sample from the population $(X,\Theta)$. In this section, we will describe how to estimate the unknown functions $r_k(\cdot)$, $\mu(\cdot)$ and $\sigma^2(\cdot)$ given the random sample. Section \ref{sec:param} describes the parameter estimation when certain parametric forms of $r_k(\cdot)$, $\mu(\cdot)$ and $\sigma^2(\cdot)$ are assumed, and Section \ref{sec:nonparam} contains the non-parametric estimation. Section \ref{sec:init} discusses how to achieve the good initial solutions for both of the cases. 

\subsection{Likelihood Maximization For Parametric Case} \label{sec:param}
Assume the parametric forms $\mu(\V{x}) = \V{B}(\V{x})' \V{\beta}$, and $\sigma^2(\V{x}) = \sigma^2$, where $\V{B}(\V{x})$ be the $q$-dimensional vector of nonlinear basis function values $B_j(\V{x})$'s, and $\V{\beta}$ is an $q$ dimensional vector of unknown coefficients. The log likelihood function becomes a function of $\V{\beta}$ and $\sigma^2$. 
\begin{equation}
\mathcal{L}(\V{\beta}, \sigma^2) = \sum_{i=1}^n \log\left\{\sum_{k=-K}^K f(\theta_i|\V{B}(\V{x})' \V{\beta} - (2k+1)\pi, \sigma^2)\right\}.
\end{equation}
The log likelihood for the standard Gaussian mixture model can be easily maximized using the standard EM algorithm:
\begin{description}
\item[Initialize:] Get the initial estimates of $\boldsymbol{\beta}$, $\sigma^{2}$ and $r_k$ using Section \ref{sec:init}.
\item[E-Step:] Compute
\begin{equation}
\psi_{i,k} = \dfrac{f(\theta_i|\V{B}(\V{x}_i)' \V{\beta} - (2k+1)\pi, \sigma^2)}{\sum_{j=-K}^{K} f(\theta_i|\V{B}(\V{x}_i)' \V{\beta} - (2j+1)\pi, \sigma^2)}.
\end{equation}
\item[M-Step:] Update the parameter estimation
\begin{equation*}
\begin{split}
& \boldsymbol{\beta} = \dfrac{\sum_{i=1}^{n}\sum_{k=-K}^{K}\psi_{i,k}(\theta_i + (2k+1)\pi)\V{B}(\V{x}_i)}{\sum_{i=1}^{n}\sum_{k=-K}^{K}\psi_{i,k}\V{B}(\V{x}_i)'\V{B}(\V{x}_i)}
\\*
& \sigma^{2} = \dfrac{\sum_{i=1}^{n}\sum_{k=-K}^{K}\psi_{i,k}(\theta_i-\V{B}(\V{x}_i)'\boldsymbol{\beta} + (2k+1)\pi )^2}{\sum_{i=1}^{n}\sum_{k=-K}^{K}\psi_{i,k}}
.
\end{split}
\end{equation*}
\end{description}

The EM algorithm requires the predetermined number of mixture components $K$. The choice of the mixture components for a finite Gaussian mixture model has been extensively studied. The comprehensive review can be found at \citet{mclachlan2014number}. In this paper, we use the Bayesian information criterion approach, 
\begin{equation*}
-2 \mathcal{L}_K + \log(n) df_K,
\end{equation*}
where $\mathcal{L}_K$ and $df_K$ are the maximum log likelihood and the degree of freedom for a choice $K$ respectively. The degree of freedom is equal to the total number of parameters,
\begin{equation*}
df_K = q + 2K+1.
\end{equation*}
\noindent\textit{\textbf{Remark.} The posterior estimation of the parameters can be also easily performed by a Gibbs sampler when conjugate priors are used, $\V{\beta} \sim \mathcal{N}\left(0, \sigma_0^2 \M I \right)$, $\sigma^2 \sim \mbox{Inv-Gamma}\left(\alpha,\lambda\right)$, and $\V{r} \sim \mbox{Dirichlet}\left(\V{\gamma}\right)$, with hyper priors  $\sigma_0^2 \sim \mbox{Inv-Gamma}\left(\alpha_0 , \lambda_0\right)$. We will skip the description of the Gibbs sampler since it is already well discussed in the literature \citep{viele2002modeling}.}

\subsection{Likelihood Maximization For Nonparametric Case} \label{sec:nonparam}
Assume $\mu(\V{x})$ and $\sigma^2(\V{x})$ are non-parametric functions. The log likelihood function for the random sample is 
\begin{equation} 
\mathcal{L} = \sum_{i=1}^n \log \left\{\sum_{k=-K}^K f(\theta_i|\mu(\V{x}_i) - (2k+1)\pi, \sigma^2(\V{x}_i))\right\}.
\end{equation}
\citet{huang2013nonparametric} studied a more general form of the log likelihood,
\begin{equation} 
\mathcal{L}^{H} = \sum_{i=1}^n \log \left\{\sum_{k=-K}^K f(\theta_i|\mu_k(\V{x}_i), \sigma_k^2(\V{x}_i))\right\}.
\end{equation}
Our log likelihood is its special case with $\mu_k(\V{x}_i) = \mu(\V{x}_i) - (2k+1)\pi$ and $\sigma_k^2(\V{x}_i) = \sigma^2(\V{x}_i)$. We employ \citet{huang2013nonparametric} to estimate non-parametric functions $\mu(\cdot)$ and $\sigma^2(\cdot)$. The approach takes the kernel regression approach to approximate the the non-parametric functions. In the kernel regression, a finite number of grid points $\{\V{x}^{(j)}: j = 1,...,J\}$ are pre-selected, and the non-parametric functions are locally approximated at each $\V{x}^{(j)}$ by local constants, i.e., 
\begin{equation*}
\mu_j \approx \mu(\V{x}^{(j)}), \mbox{ and } \sigma^2_j \approx \sigma^2(\V{x}^{(j)}).
\end{equation*}
Then, the non-parametric function values at an arbitrary location $\V{x}$ are achieved by interpolating the local constants; we used a linear interpolation for numerical examples. For the maximum likelihood estimates of the local constants, an EM algorithm is proposed. Let $C_h(\cdot) = h^{-1}C(\cdot/h)$ denote the kernel function with bandwidth $h$. The local log likelihood function at $\V{x}^{(j)}$ is define as
\begin{equation} 
\mathcal{L}_j = \sum_{i=1}^n \log\left\{\sum_{k=-K}^K f(\theta_i|\mu_j - (2k+1)\pi, \sigma^2_j) \right\} C_h(||\V{x}_i - \V{x}^{(j)}||_2).
\end{equation}
The local log likelihood is maximized to estimate the local constants $\mu_j$ and $\sigma^2_j$ using the following EM steps:
\begin{description}
\item[Initialize:] Get the initial estimates of $\mu(\cdot)$ and $\sigma^2(\cdot)$ using Section \ref{sec:init}. Use the estimates to evaluate $\mu(\V{x}_i)$, and $\sigma^2(\V{x}_i)$. 
\item[E-step:] Compute
\begin{equation}
\psi_{i,k} = \dfrac{f(\theta_i| \mu(\V{x}_i) - (2k+1)\pi, \sigma^2(\V{x}_i))}{\sum_{k'=-K}^{K} f(\theta_i| \mu(\V{x}_i) - (2k'+1)\pi, \sigma^2(\V{x}_i))}.
\end{equation}
\item[M-step:] Update the estimation of the local constants 
\begin{equation*}
\begin{split}
& \mu_j = \dfrac{\sum_{i=1}^{n}\sum_{k=-K}^K \psi_{i,k} C_h(||\V{x}_i - \V{x}^{(j)}||_2) (\theta_i+ (2k+1)\pi) }{\sum_{i=1}^{n}\sum_{k=1}^K \psi_{i,k} C_h(||\V{x}_i - \V{x}^{(j)}||_2)}.
\\*
& \sigma_j^{2} = \dfrac{\sum_{i=1}^{n}\sum_{k=-K}^{K}\psi_{i,k} C_h(||\V{x}_i - \V{x}^{(j)}||_2)(\theta_i - \mu_j + (2k+1)\pi )^2}{\sum_{i=1}^{n}\sum_{k=1}^{K}\psi_{i,k}C_h(||\V{x}_i - \V{x}^{(j)}||_2)}.
\end{split}
\end{equation*}
Update the values of $\mu(\V{x}_i)$, and $\sigma^2(\V{x}_i)$ by interpolating the estimated $\mu_j$'s and $\sigma^2_j$'s respectively.
\end{description}
The convergence of the EM algorithm were studied in \citet[Theorem 3]{huang2013nonparametric}.

There are three tuning parameters, the number of mixture components $K$, the bandwidth parameter $h$ and the locations and number of grid points $\V{x}^{(j)}$ for local regression. The grid locations can be selected among the observations $\V{x}_i$. In particular, this is more efficient when the input dimension $p$ is high, because the uniform selection of the grid locations over a high dimensional space produces a huge number of the grid locations. Regarding the selection of $K$ and $h$, we follow \citet{huang2013nonparametric}, which first selects $K$ that minimizes the BIC for certain ranges of values of $K$ and $h$ and then chooses $h$ using a multi-fold cross validation.     

\subsection{Initialization of Parameters} \label{sec:init}
In this section we will discuss how to achieve good initial estimates of the model parameters that are necessary to initiate the EM methods described in Sections \ref{sec:param} and \ref{sec:nonparam}. We first estimate $Z_i$ and use them to estimate the model parameters. Note that $Z_i$ represents a class label, and we use a clustering algorithm to estimate the variable. First apply a clustering algorithm to find the disparate clusters of $\{(\V{x}_i, \theta_i)\}$. For all of our numerical examples, we applied the density-based clustering algorithm \citep{ester1996density}. Suppose that $J_k$ is the set of $i$'s that index the elements belonging to the $k$th cluster. Please note that the number of the clusters identified by the clustering algorithm is not necessarily same as the number of the mixture components in AGMM. Therefore, we use a different symbol $\tilde{K}$ to denote the number of the clusters. We will map the $k$th cluster to the AGMM mixture component number $s_k$ as follows:
\begin{equation*}
Z_i = s_k \mbox{ for } i \in J_k \quad ,\forall k \in \{1, \ldots, \tilde{K}\}.
\end{equation*}
We first assign $s_1 = 0$ as a baseline for the first cluster, i.e.,
\begin{equation*}
Z_i = 0 \mbox{ for } i \in J_1,
\end{equation*}
and sequentially assign $Z_i$'s for the other clusters as follows: for $k \in \{2, \ldots, \tilde{K}\}$, find $k^*$ that 
\begin{equation*}
k^* = \arg\min\{ d_H(J_k, J_{k'}); k' = 1,...,k-1. \},
\end{equation*}
where $d_H(J_k, J_{k'}) = \min_{i \in J_k, j \in J_{k'}} ||\V{x}_i -\V{x}_j||_2$ quantifies the distance in between cluster $k$ and cluster $k'$. Assign 
\begin{equation*}
s_k = s_{k^*} + \mbox{round}\left( \frac{\Theta_{j_{k^*}}-\Theta_{i_k}}{2\pi} \right),
\end{equation*}
where $(i_k, j_{k^*}) = \arg\min_{i \in J_k, j \in J_{k^*}} ||\V{x}_i -\V{x}_j||_2$. Once the initial assignments on $Z_i$ are completed, one may run either the M-step of Section \ref{sec:param} or the M-step of Section \ref{sec:nonparam} with $\psi_{i,k} = 1$ only if $Z_i = k$.

\section{Numerical Examples} \label{sec:example}
In this section, the proposed AGMM will be applied for five different scenarios with varying concentration parameters, and the results will be analyzed and compared with the non-parametric smoothing \citep{di2013non} and the projected linear model \citep{hernandez2017general}. For the comparison purpose, we used the results from the parametric AGMM model (described in Section \ref{sec:param}); the comparison of the parametric AGMM and the non-parametric AGMM (described in Section \ref{sec:nonparam}) will be separately presented in Section \ref{sec:AGMMcomp}. For the parametric AGMM, we used the B-spline basis functions of degree three with evenly spaced knot locations for $\V{B}(\V{x})$, and the number of knots was determined by the 5-fold cross validation. The parameter estimation of the parametric AGMM was performed using the Gibbs sampler to make the fair comparison of computation times in between AGMM and the Gibbs sampler of the projected linear model. For the projected linear model, we tried all of the linear, quadratic, and cubic models that were used to represent its mean function in their paper, and only presented the best result for each example. We used the Gibbs sampler proposed in the original paper to estimate the projected linear model, and 30,000 Gibbs sampling steps were taken while the first 10,000 burn-in samples were not used. For the non-parametric smoothing, we used a triangular kernel, and the 5-fold cross validation was used to select the kernel parameter. All computations were performed in a desktop computer with Intel Core i7-6600U CPU and 16 GB RAM.

For this numerical comparison, we used the mean circular error (MCE) to measure the test errors of the tested algorithms to the ground truth. We define $|\sin\left(\frac{\theta - \hat{\theta}}{2}\right)|$ as the angular distance between $\theta$ and $\hat{\theta}$, which ranges in $[0, 1]$; it is equal to one when the two angles have the maximum angular difference (i.e. $\pi$) on a circle, and it approaches to zero as the difference approaches to zero. The MCE is the arithmetic average of the distance measure over a test dataset,
\begin{equation*}
\mbox{MCE} = \frac{1}{T}\sum_{i=1}^T \left|\sin\left(\frac{\theta_i - \hat{\theta}_i}{2}\right)\right|,
\end{equation*}
where $\theta_i$ is the ground truth at the $i$th testing location, and $\hat{\theta}_i$ is the estimated output at the same testing location. 

\subsection{Example 1: Synthetic von Mises - Linear Mean}
We randomly sampled 160 data points from $X \sim \Unif(-1, 1)$ and $\Theta | X=x$ following the von Mises distribution $VM(\mu(x) ,\kappa)$ with mean $\mu(x) = 0.1 + 5(x-0.5)$. The concentration parameter $\kappa$ was varied over different values of $1, 2, 4$ and $8$. A larger $\kappa$ implies random samples have larger variations around the mean. The $R$ package \texttt{circular} was used to generate the data. The estimated mean functions for AGMM, non-parametric smoothing \citep{di2013non} and the projected linear model \citep{hernandez2017general} were compared with the ground truth $\mu(x) = 0.1 + 5(x-0.5)$ at 200 randomly sampled locations. 

\begin{figure}[t]
	\includegraphics[width = 0.7\linewidth]{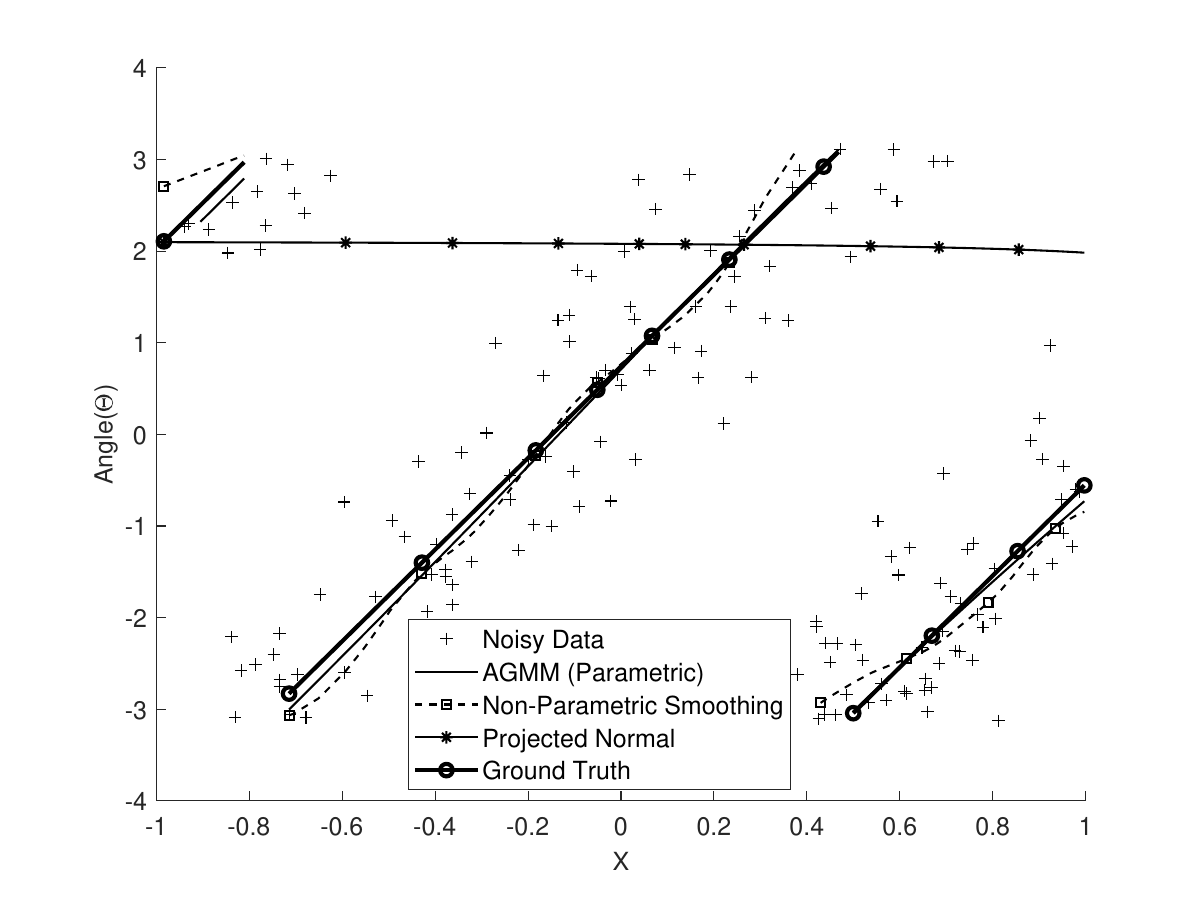}
	\centering
	\caption{Comparison of the estimated mean functions for Example 1 when $\kappa = 2$.}
	\label{fig:example1}
\end{figure}

Figure \ref{fig:example1} illustrates the estimated mean functions with the ground truth when $\kappa = 2$. In the figure, the ground-truth mean and the mean estimates are continuous in $x$ but they appear discontinuous because the ranges of the means are out of $[-\pi, \pi]$, and the means are wrapped around a sphere $[-\pi, \pi]$. The AGMM and the non-parametric smoothing estimated the mean closely to the ground truth, while the projected linear model significantly deviated from the ground truth. The mean estimate of the non-parametric smoothing is wavy around the ground truth. The AGMM achieved the lowest MCE for most of the cases; see Table \ref{ErrorTable1}. 

\begin{table}[h]
	\centering
	\begin{tabular}{|c|c|c|c|}
		\hline
		Mean Circular  & AGMM  & Non-Parametric & Projected Normal \\ 
		Error (MCE) & (Parametric) & Smoothing  & Model \\ \hline
		$\kappa=1$ & 0.1255 &  0.1365 & 0.6101 \\ \hline
		$\kappa=2$ & 0.0409 & 0.0829 & 0.6159 \\ \hline
		$\kappa=4$ & 0.0663 & 0.0614 & 0.5686 \\ \hline
		$\kappa=8$ &  0.0231 &  0.0679 &  0.5537 \\ \hline
	\end{tabular} 
	\caption {Estimation Accuracy for Example 1. This table demonstrates the minimum circular errors (MCEs) of the estimated mean functions of the three compared methods to the ground truth.}
	\label{ErrorTable1}
\end{table}

\subsection{Example 2: Synthetic von Mises - Nonlinear Mean} 
In this example we followed \citet{nunez2011bayesian} to generate the dataset used in their work. It consists of 80 random draws from $X \sim \Unif(-1,1)$ and $\Theta|X=x$ following the von Mises distribution $VM(\mu(x),\kappa)$ with mean $\mu(x) = 0.1 + \arctan(5x)$. The concentration parameter $\kappa$ was varied over different values of $1, 2, 4$ and $8$. The $R$ package \texttt{circular} was used to generate the data. The estimated mean functions for the three methods were compared with the ground truth $\mu = 0.1 + \arctan(5x)$ at 200 randomly selected locations.

\begin{figure}[t]
	\includegraphics[width = \linewidth]{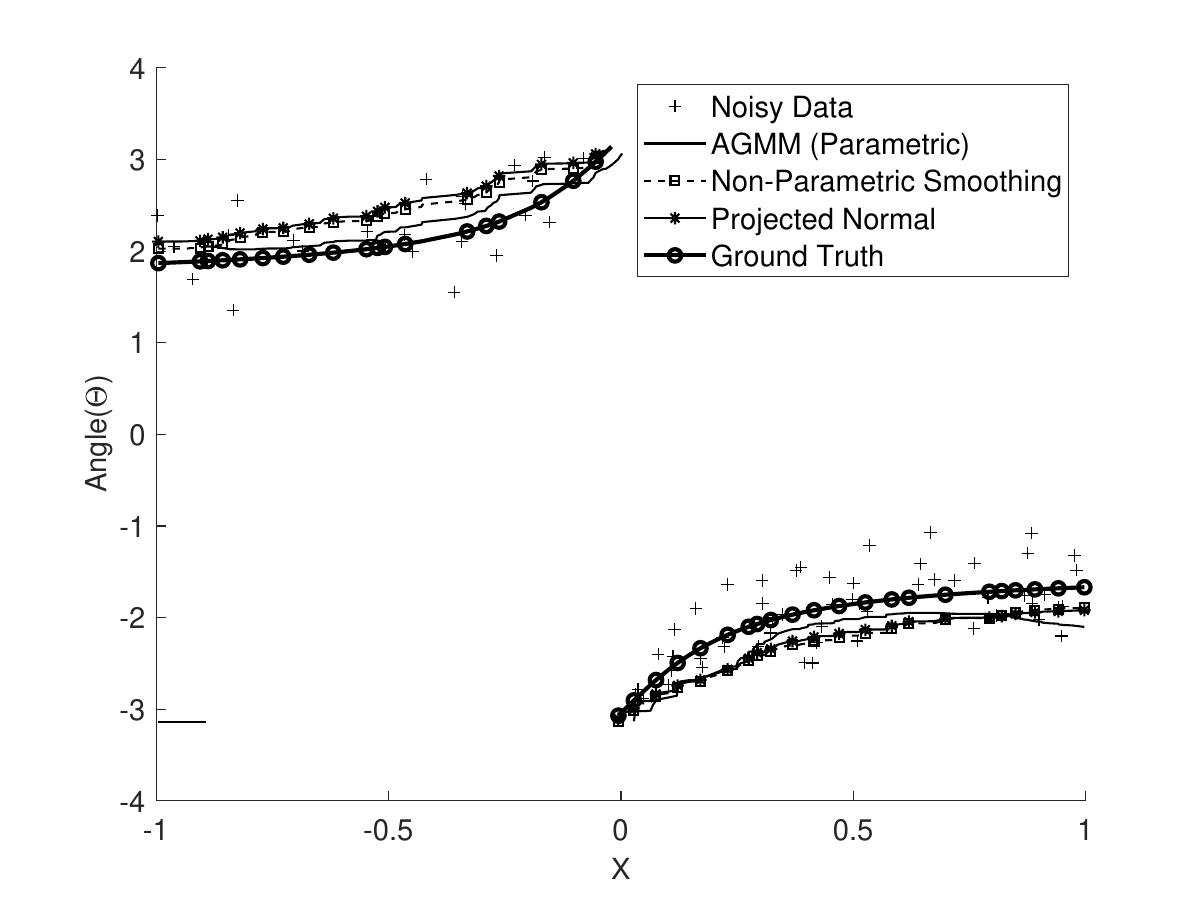}
	\centering
	\caption{Comparison of the estimated mean functions for Example 2 with $\kappa = 2$.}
	\label{fig:example2}
\end{figure}

Figure \ref{fig:example2} shows the estimated mean functions for $\kappa=2$. The estimates of all compared methods were within a reasonable range to the ground truth. Table \ref{ErrorTable2} compares the MCE values of the three methods. The non-parametric smoothing outperformed the two other methods for the lowest concentration case, while the AGMM was the best performer for the other three cases.  

\begin{table}[h]
	\centering
	\begin{tabular}{|c|c|c|c|}
		\hline
		Mean Circular  & AGMM  & Non-Parametric & Projected Normal \\ 
		Error (MCE) & (Parametric) & Smoothing  & Model \\ \hline
		$\kappa=1$ & 0.1503 &  0.0964 & 0.1025 \\ \hline
		$\kappa=2$ & 0.1224 & 0.1488 & 0.1613 \\ \hline
		$\kappa=4$ & 0.0404 & 0.0525 & 0.0704 \\ \hline
		$\kappa=8$ &  0.0416 &  0.0484 &  0.0647 \\ \hline
	\end{tabular} 
	\caption {Estimation Accuracy for Example 2. This table demonstrates the minimum circular errors (MCEs) of the estimated mean functions of the three compared methods to the ground truth.}
	\label{ErrorTable2}
\end{table}

\subsection{Example 3: Synthetic Projected Normal - Linear Mean}
We randomly sampled 300 data points from $X \sim \Unif(-1, 1)$ and $\Theta | X=x$ following the projected normal distribution $PN(\V{\mu}(x), \sigma^2\M{I})$ with 
\begin{equation}
\V{\mu}(x) = \left[\begin{array}{c} 1 + 2(x-2) \\ -4x \end{array}\right], 
\end{equation}
and the variance parameter $\sigma^2 \in \{1, 4, 9\}$. The estimated mean functions for the three methods were compared with the ground truth at 100 randomly selected locations.

\begin{figure}[t]
	\includegraphics[width = 0.8\linewidth]{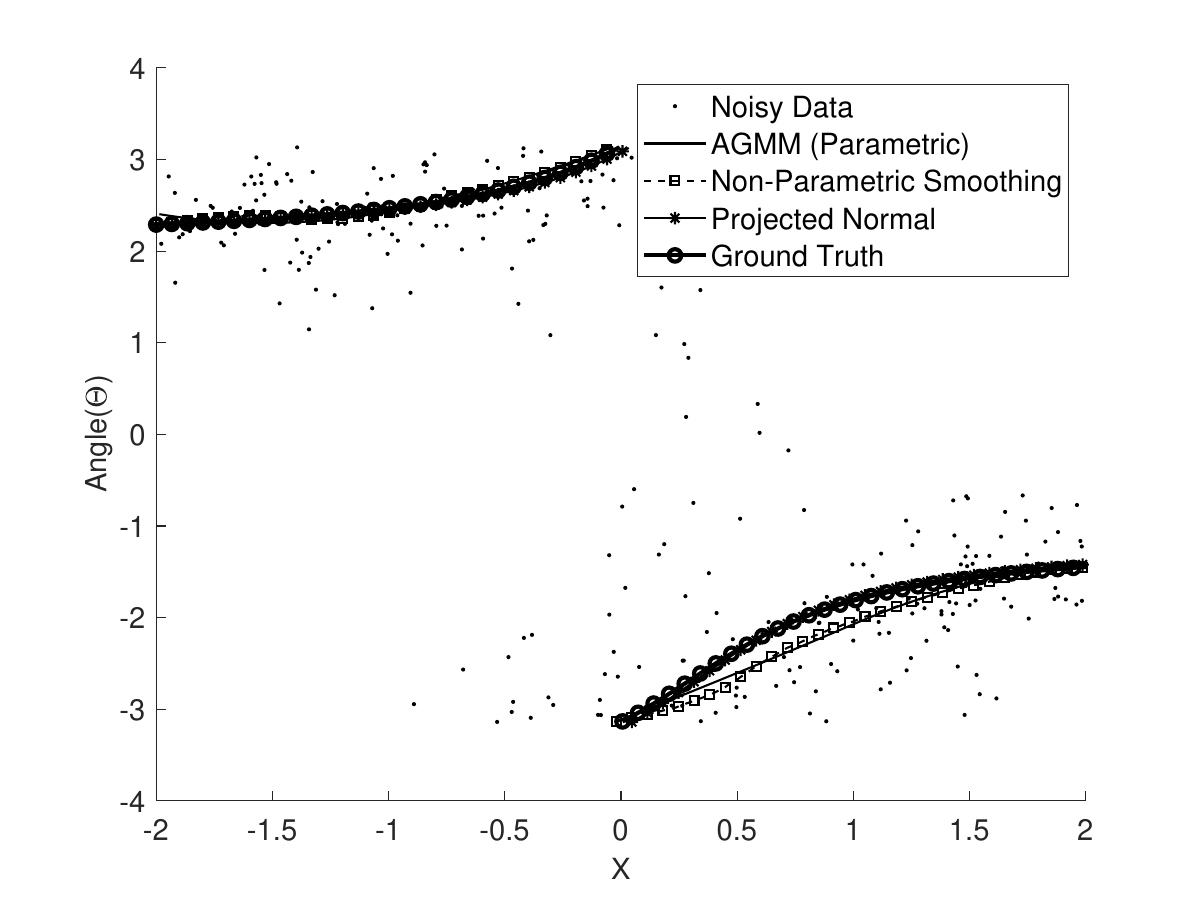}
	\centering
	\caption{Comparison of the estimated mean functions for Example 3 with $\sigma^2 = 9$.}
	\label{fig:example3}
\end{figure}

Figure \ref{fig:example3} shows the estimated mean functions for $\sigma^2=9$. The estimated mean functions of the projected normal method is the best performer for this example, running running very closely to the ground-truth, while the AGMM and the non-parametric smoothing method are comparable with their mean predictions being a little more deviated from the ground truth. Table \ref{ErrorTable3} quantitatively reflects this with the MCE values. The projected normal method outperformed the other two methods.    

\begin{table}[h]
	\centering
	\begin{tabular}{|c|c|c|c|}
		\hline
		Mean Circular  & AGMM  & Non-Parametric & Projected Normal \\ 
		Error (MCE) & (Parametric) & Smoothing  & Model \\ \hline
		$\sigma^2=1$ & 0.0190 &  0.0192 & 0.0096 \\ \hline
		$\sigma^2=4$ & 0.0273 & 0.0134 & 0.0361 \\ \hline
		$\sigma^2=9$ & 0.0493 & 0.0498 & 0.0141 \\ \hline
	\end{tabular} 
	\caption {Estimation Accuracy for Example 3. This table demonstrates the minimum circular errors (MCEs) of the estimated mean functions of the three compared methods to the ground truth.}
	\label{ErrorTable3}
\end{table}

\subsection{Example 4: Synthetic Projected Normal - Nonlinear Mean}
We randomly sampled 160 data points from $X \sim \Unif(-1, 1)$ and $\Theta | X=x$ following the projected normal distribution $PN(\V{\mu}(x), \sigma^2\M{I})$ with 
\begin{equation}
\V{\mu}(x) = \left[\begin{array}{c} 4\sin(2\pi x) \\ 8\cos(2\pi x - \pi/2)+4\end{array}\right], 
\end{equation}
and the variance parameter $\sigma^2 \in \{1, 4, 9\}$. The estimated mean functions for the three methods were compared with the ground truth at 100 randomly selected locations.

\begin{figure}[t]
	\includegraphics[width = 0.8\linewidth]{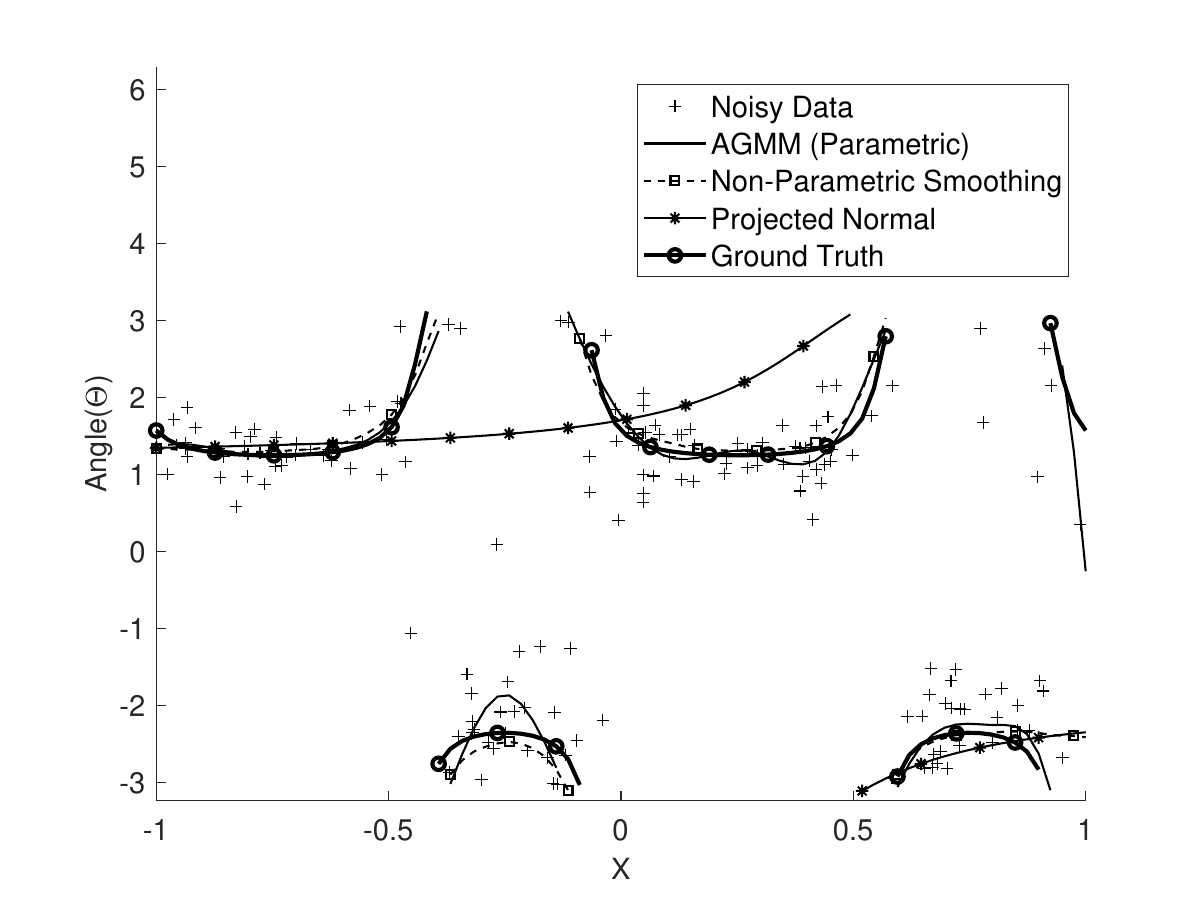}
	\centering
	\caption{Comparison of the estimated mean functions for Example 4 with $\sigma^2 = 9$.}
	\label{fig:example4}
\end{figure}

Figure \ref{fig:example4} shows the estimated mean functions for $\sigma^2=9$. The estimated mean functions of the AGMM and the non-parametric smoothing method are comparable, running closely to the ground-truth, while the projected normal method is significantly deviating from the ground-truth. Table \ref{ErrorTable4} quantitatively reflects this with the MCE values. The AGMM and the non-parametric smoothing outperformed the projected normal method.    

\begin{table}[h]
	\centering
	\begin{tabular}{|c|c|c|c|}
		\hline
		Mean Circular  & AGMM  & Non-Parametric & Projected Normal \\ 
		Error (MCE) & (Parametric) & Smoothing  & Model \\ \hline
		$\sigma^2=1$ & 0.0868 &  0.0695 & 0.3948 \\ \hline
		$\sigma^2=4$ & 0.0974 & 0.0698 & 0.3761 \\ \hline
		$\sigma^2=9$ & 0.0921 & 0.1004 & 0.3660 \\ \hline
	\end{tabular} 
	\caption {Estimation Accuracy for Example 4. This table demonstrates the minimum circular errors (MCEs) of the estimated mean functions of the three compared methods to the ground truth.}
	\label{ErrorTable4}
\end{table}

\subsection{Example 5: Synthetic Wrapped Normal}
\begin{figure}[t]
	\includegraphics[width = 0.8\linewidth]{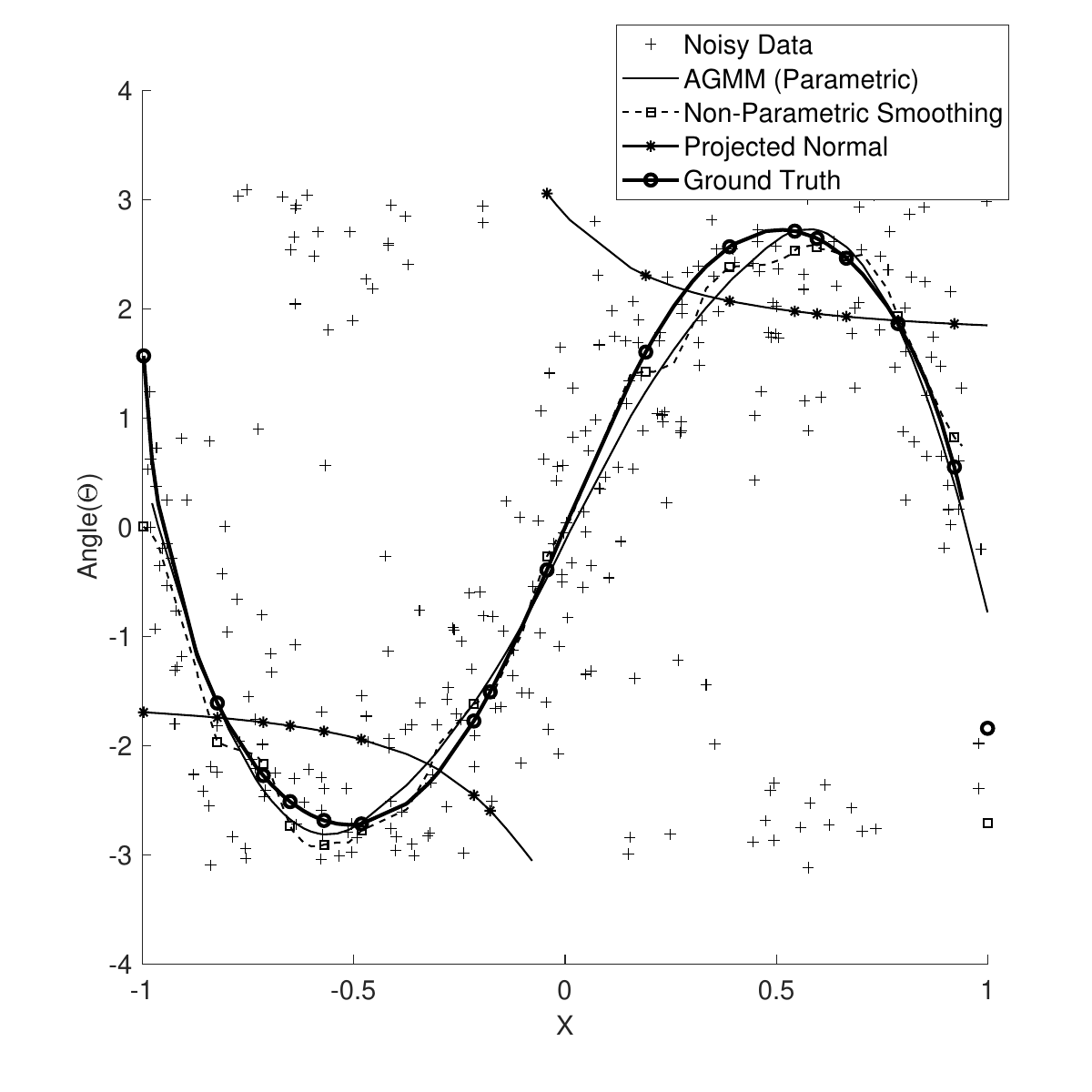}
	\centering
	\caption{Comparison of the estimated mean functions for Example 5 with $\sigma^2=1$}
	\label{fig:example5}
\end{figure}

In this example we randomly sampled 300 data points from $X \sim \Unif(-1, 1)$ and 
\begin{equation*}
Y|X=x  \sim \mathcal{N}\left(\mu(x), \sigma^2\right),
\end{equation*}
where $\mu(x) = \left(\arctan\left(2x\right) + \arcsin\left(\frac{x}{2}\right) - \arcsin\left(x\right) + \arccos\left(\frac{x}{3}\right) - \frac{\pi}{2}\right)\times 7.85 + \pi$. The noise parameter $\sigma^2$ was varied over $0.5, 0.7$ and $1.0$. When $y_i$ denotes the $i$th sample, the $i$th circular response can be achieved from $y_i$ by taking $\theta_i = (y_i \mod 2\pi) - \pi$. The mean function of $Y|X=x$ is continuous, and its range fits in $[-\pi, \pi]$. Therefore the modulo of the mean, i.e., $(E[Y|X=x] \mod 2\pi) - \pi$, should be continuous. However, due to the population variance, the sampled $y_i$ value can be out of the range $[-\pi, \pi]$. Therefore, the corresponding $\theta_i$ values in the random sample can be divided into multiple disconnected pieces as the results of the modulo operation on $y_i$. Figure \ref{fig:example5} shows the random sample split in three pieces. 

We compared the mean estimates of the three methods with the ground truth. Figure \ref{fig:example5} illustrates the comparison for the lowest concentration case that we tried, and Table \ref{ErrorTable5} summarizes the MCE performance for all tested cases. The AGMM and the non-parametric smoothing produced the mean estimates close to the ground truth, while the projected normal method produced the mean estimates significantly deviating from the ground truth. For the projected linear model, all of linear, quadratic, and cubic models for the mean function were tested, and none of the choices produced a good fit. The result with the quadratic mean model is displayed in Figure \ref{fig:example5}. 
\begin{table}[h]
	\centering
	\begin{tabular}{|c|c|c|c|}
		\hline
		Mean Circular  & AGMM  & Non-Parametric & Projected Normal \\ 
		Error (MCE) & (Parametric) & Smoothing  & Model \\ \hline
		$\sigma^2=0.5$ & 0.0447 &  0.0584 & 0.3568\\ \hline
		$\sigma^2=0.7$ & 0.0677 & 0.0736 & 0.3812\\ \hline
		$\sigma^2=1.0$ & 0.0810 & 0.0930 & 0.3888 \\ \hline
	\end{tabular} 
	\caption {Estimation Accuracy for Example 5. This table demonstrates the minimum circular errors (MCEs) of the estimated mean functions of the three compared methods to the ground truth.}
	\label{ErrorTable5}
\end{table}

\subsection{Parametric AGMM vs. Non-parametric AGMM} \label{sec:AGMMcomp}
In this section, we compare the parametric AGMM model (described in Section \ref{sec:param}) and the non-parametric AGMM model (described in Section \ref{sec:nonparam}) for Example 1 and Example 5. For the non-parametric AGMM, all $\V{x}_i$'s of the random sample are chosen as the grid locations $\V{x}^{(j)}$'s to form non-parametric functions. We fixed $h = 0.01$ and $K$ was simply chosen using the BIC. The mean and variance estimates are achieved using the formula described in the M-step of Section \ref{sec:nonparam}. For the parametric AGMM, we ran 30,000 Gibbs sampling iterations with the first 10,000 as burn-in samples. The mean and variance estimates are achieved taking the averages of the corresponding sampled values taken after the burn-in period. Taking the variance estimates does require a very little marginal time over the time for taking only the mean estimates for both of the parametric and nonparametric AGMM. 

\begin{figure}[p]
	\includegraphics[width = \linewidth]{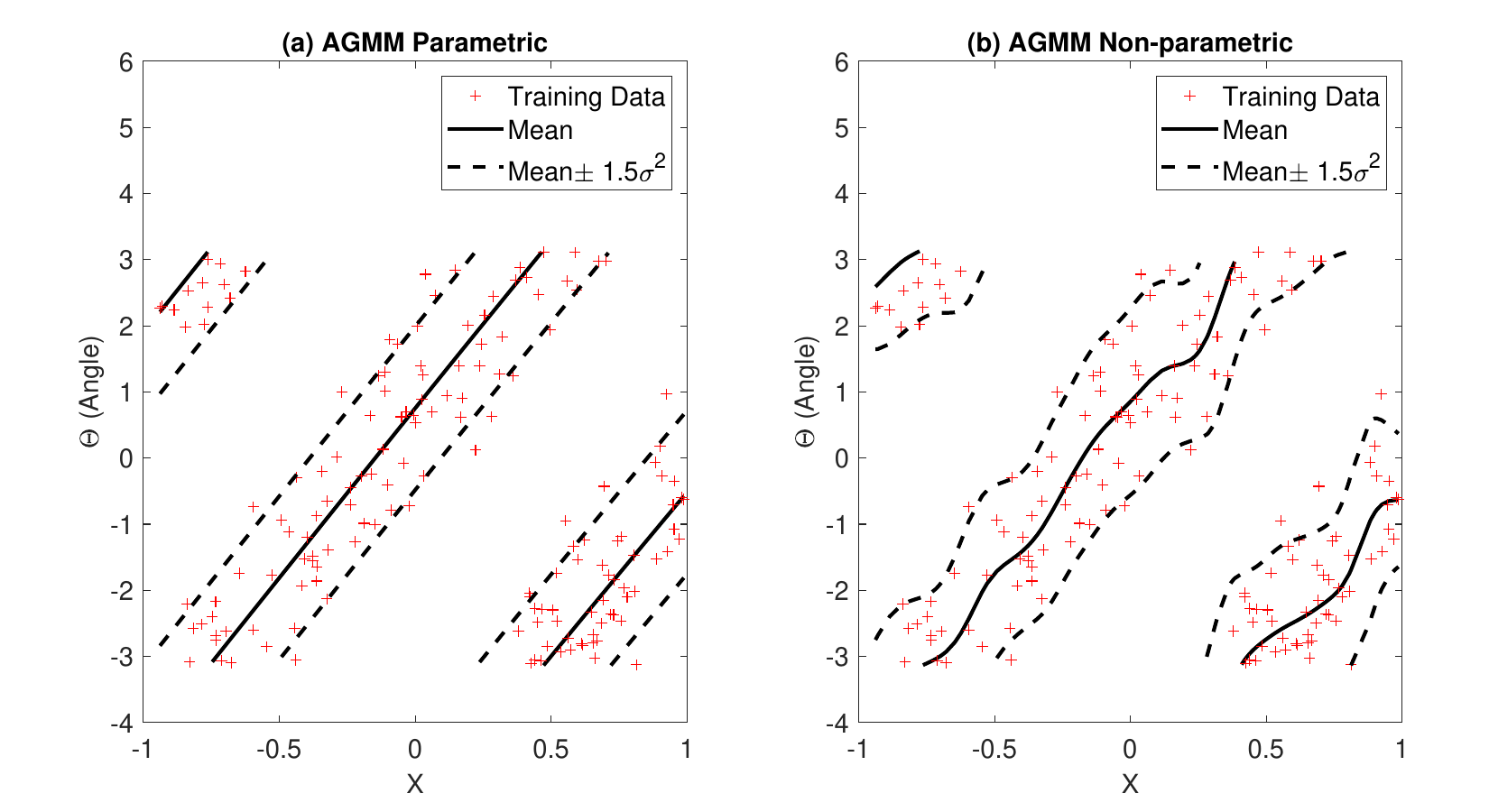}
	\centering
	\caption{Comparison of Parametric AGMM and Non-parametric AGMM for Example 1.}
	\label{fig:example1-comp}
	\includegraphics[width = \linewidth]{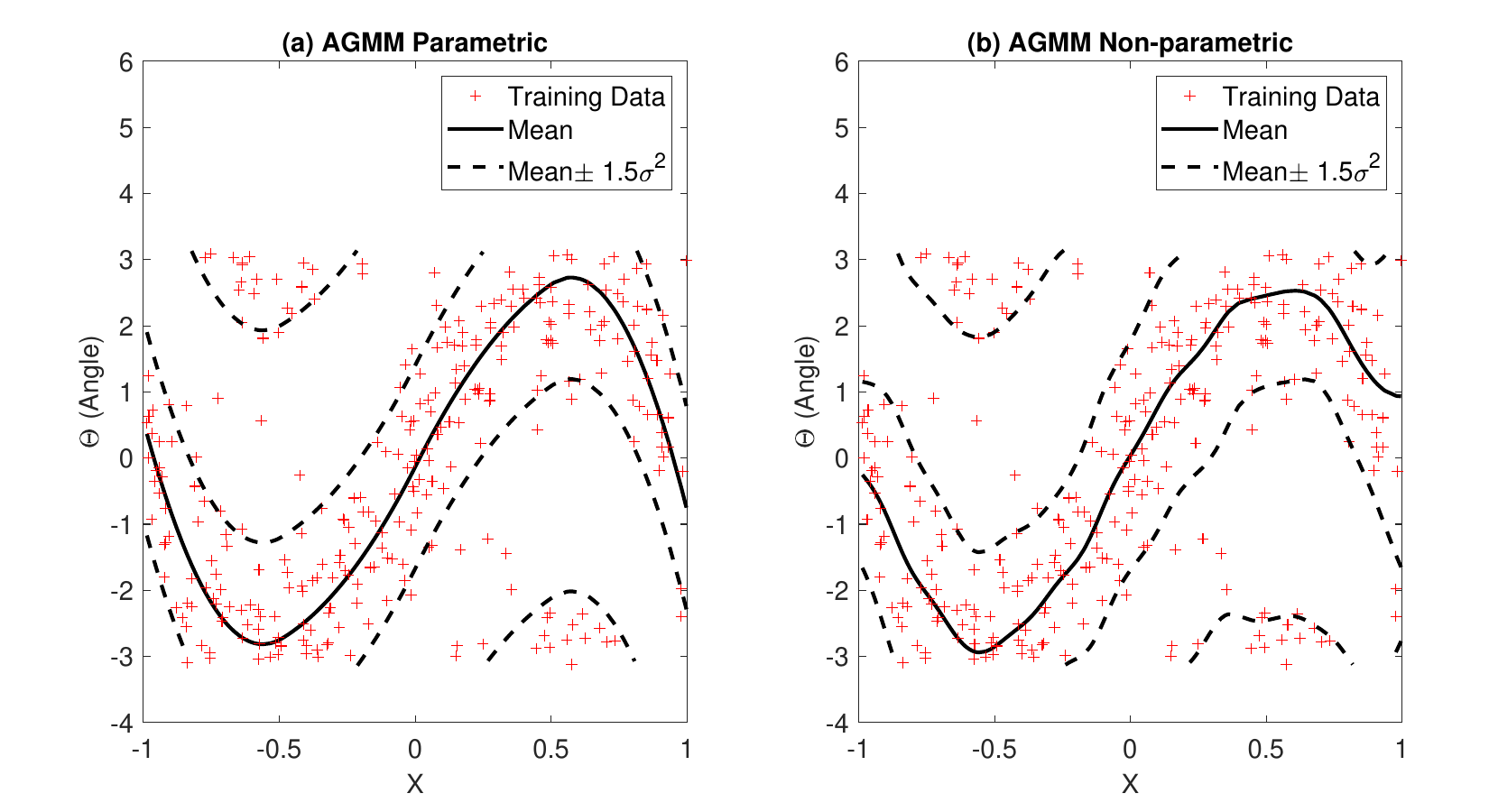}
	\centering
	\caption{Comparison of Parametric AGMM and Non-parametric AGMM for Example 5.}
	\label{fig:example5-comp}
\end{figure}

The mean estimates and the variance estimates were compared. Figures \ref{fig:example1-comp} and \ref{fig:example5-comp} show the comparison results. The mean estimates are comparable to each other, while the non-parametric AGMM is prone to overestimating the variance $\sigma^2(\cdot)$. For quantitative comparison, we compute the mean circular errors of the mean estimates and the mean square errors of the variance estimates. Table \ref{ErrorTable_AGMM} shows the errors against the ground truths. For Example 1, the ground truth is $\sigma^2 = 1 - I_1(2)/I_0(2)$, where $I_{\nu}(k)$ is the modified Bessel function of the first kind. For Example 5, the ground truth is equal to the simulation input $\sigma^2 = 1$.  The mean square errors shown in the table are the average square distance of the estimated $\sigma^2(\cdot)$ to the ground truth. 

On the other hand, the non-parametric AGMM's EM iterations converged very fast, 4 iterations for Example 1 and 62 iterations for Example 5. The total computation times were 0.59 seconds for Example 1 and 8 seconds for Example 5. For the parametric AGMM, the computation times were comparable when the EM algorithm is applied for parameter estimation; the computation times were much longer when the Gibbs sampling is used with 30,000 samples.  

\begin{table}[h]
		\centering
	\begin{tabular}{|c|c|c|}
		\hline
		MCE for $\mu(\cdot)$ & Parametric AGMM  & Non-parametric AGMM \\ 
	    (MSE for $\sigma^2(\cdot)$) & Model & Model \\ \hline
		Example 1 & 0.0409 & 0.0273 \\ 
		          & (0.1420)        & (0.2014) \\ \hline
		Example 5 & 0.0810 & 0.0737 \\ 
		& (0.5593)        & (0.8133) \\ \hline
	\end{tabular} 
	\caption {Estimation Accuracy of Parametric AGMM and Non-parametric AGMM.}
	\label{ErrorTable_AGMM}
\end{table}

\subsection{Computational Aspect}
Besides the estimation accuracy, the computation time is another important factor to be considered. In this section, we summarize the computation efficiency of the three compared methods for the five simulation cases. 

Table \ref{TimeTable} contains the total computation times of the three methods for the five simulated examples. The non-parametric smoothing does not involve any sampling, so it is fastest. Parameter tuning is the most significant part of its computation, which involves a 5-fold cross validation for choosing one kernel parameter in the triangular kernel. The computation times of the AGMM are the times for 30,000 Gibbs sampling steps, which were not very long. This is because all sampling steps with the Gibbs sampler of the AGMM are as simple as sampling from standard distributions such as normal, beta and inverse gamma distributions.

The computation times of the projected normal model were dependent on what sampling approaches were used. For example, the sampling with the Metropolis-Hastings sampling steps within a Gibbs sampler \citep{nunez2011bayesian} took a huge amount of time, e.g. 1,343,364 seconds for Example 1. However, the slice sampling approach \citep{hernandez2017general} took a computation time comparable or faster than the parametric AGMM model. However, the nonparametric version of the AGMM does not require any sampling steps, having the EM iterations instead, which is a lot more faster than the projected normal model. The computation times are shown in Table \ref{TimeTable}.

\begin{table}[h]
		\centering
	\begin{tabular}{|c|c|c|c|c|}
		\hline
		Computation Time &AGMM & AGMM & Non-Parametric  & Projected Normal   \\ 
		(Unit: seconds) & (Parametric) & (Nonparametric) &Smoothing & Model \\ \hline
		Example 1 &  30.977 Sec & 	0.59 Sec & 0.222 Sec & 26.006 Sec \\
		$n=160$ & 30,000 Iter &   		& 5-Fold CV &  30,000 Iter \\ \hline
		Example 2 & 15.833 Sec & 1.36 Sec& 0.115 Sec & 10.156 Sec \\ 
		$n=80$ & 30,000 Iter & 			& 5-Fold CV & 30,000 Iter \\ \hline
		Example 3 & 84.679 Sec & 0.90 Sec  & 1.373 Sec & 51.5535 Sec \\ 
		$n=300$& 30,000 Iter & 			& 5-Fold CV & 30,000 Iter \\ \hline
		Example 4 &  31.829 Sec & 1.51 Sec & 0.155 sec  & 26.3590 Sec \\ 
		$n=160$ & 30,000 Iter &       & 5-Fold CV  &  30,000 Iter \\ \hline
		Example 5 & 60.492 Sec & 8.01 Sec &0.501  & 45.7296 Sec \\ 
		$n=300$ & 30,000 Iter &  & 5-Fold CV  & 30,000 Iter  \\ \hline
	\end{tabular} 
	\caption {Total computations times. This table aims to provide a summary of computation time of all three methods on different examples. The projected normal model has considerably high computation time compared to the two other methods.}
	\label{TimeTable}
\end{table}

\begin{figure}
	\includegraphics[width = 0.7\linewidth]{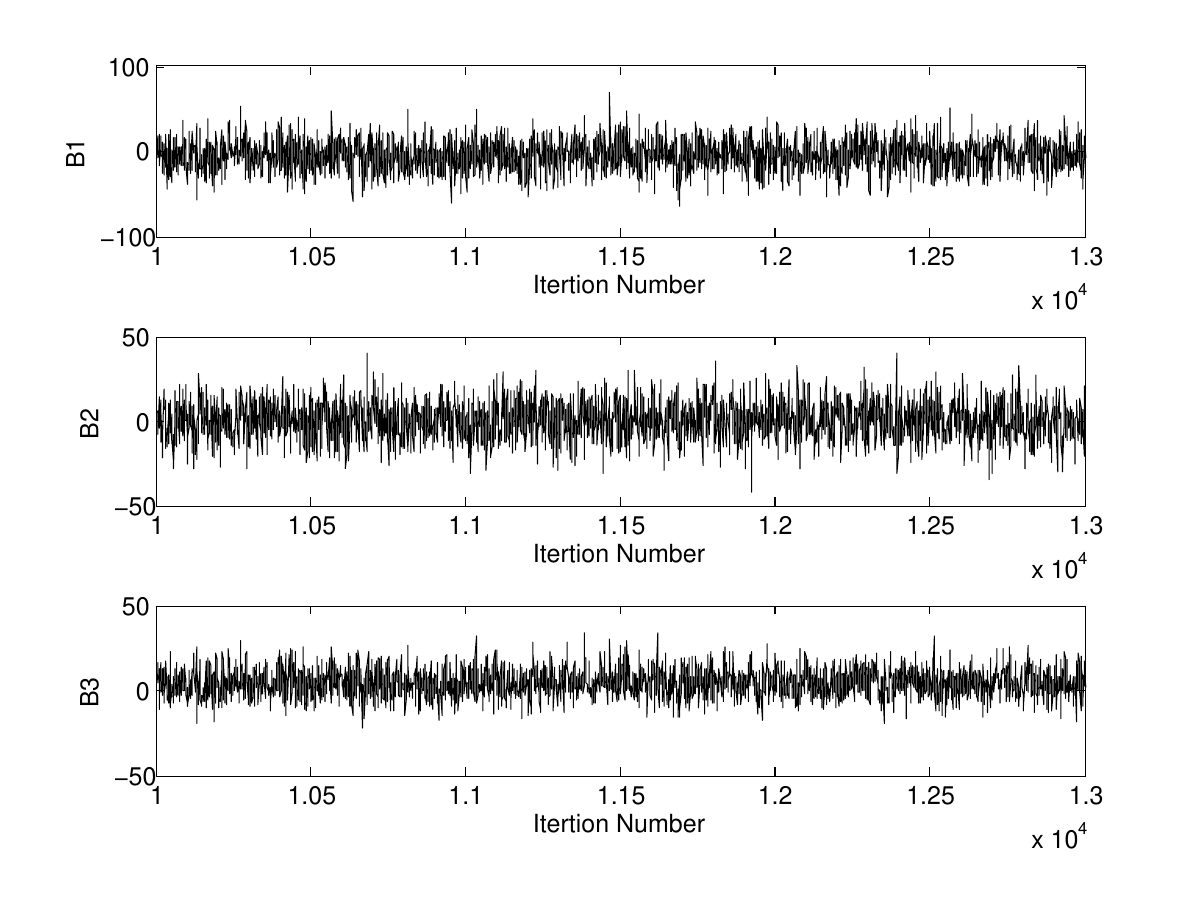}
	\centering
	\caption{ This figure shows a mixing plot from 3000 Gibbs samples of three mean parameters of the projected linear model after the first 10,000 burn-in samples.}
	\label{fig:MixingPlot1}
\end{figure}

Another practical issue with the projected normal model is that the Gibbs samplings are being kept in some bad local optima. For Example 5, we looked at the sampling results after the burn-in period of 10,000. The samples keep varying within a certain range, which correspond to bad mean estimates. Figure \ref{fig:MixingPlot1} shows the mixing plots for the mean parameters for a quadratic mean model $\mu = \beta_{1} + \beta_{2}X + \beta_{3}X^{2}$. With the sampling range, the estimated mean function is not a good fit to the ground truth as we previously saw in Figure \ref{fig:example4}.

On the other hand, the non-parametric smoothing worked well for all of the five cases with a very short computation time. Its mean estimates were very comparable to those of the AGMM. However, the non-parametric smoothing only yields the mean estimates, while the AGMM provides both of the mean and variance estimates as illustrated in Figures \ref{fig:example1-comp} and \ref{fig:example5-comp}. 

\section{Real Application: Wind Directions} \label{sec:app}
\begin{figure}[p]
	\includegraphics[width = \linewidth]{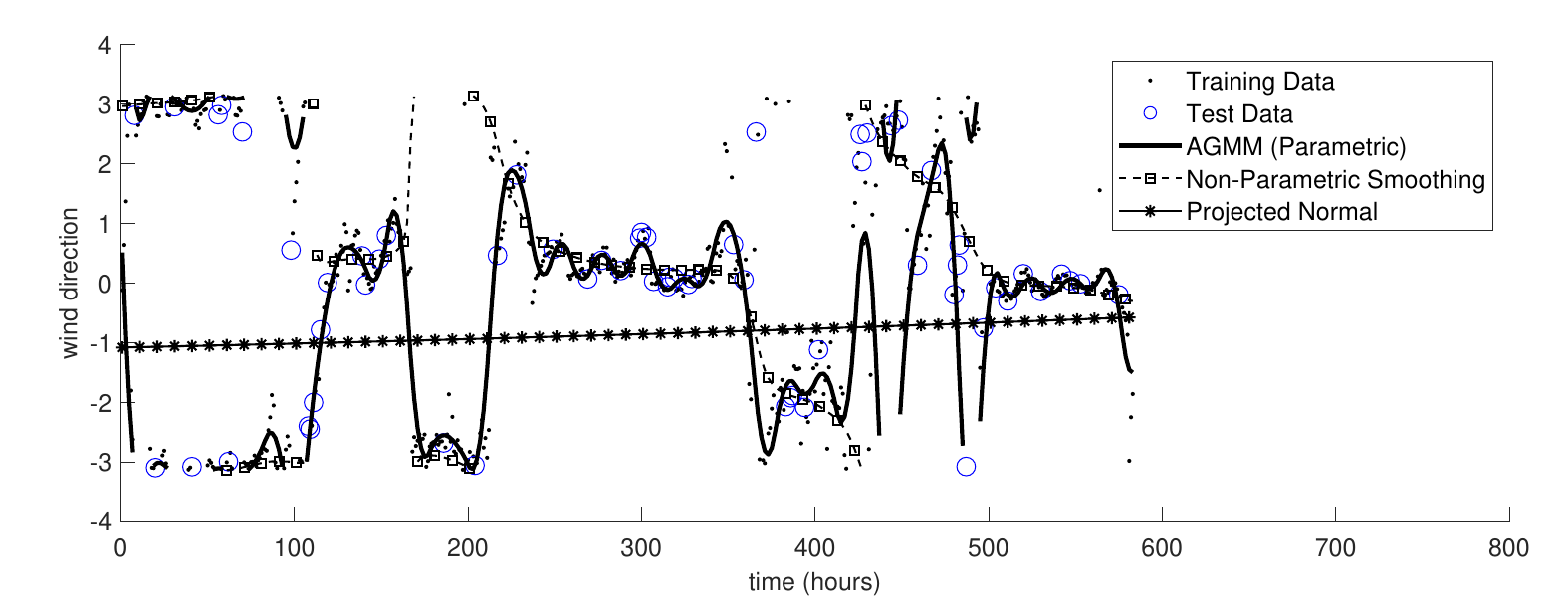}
	\includegraphics[width = \linewidth]{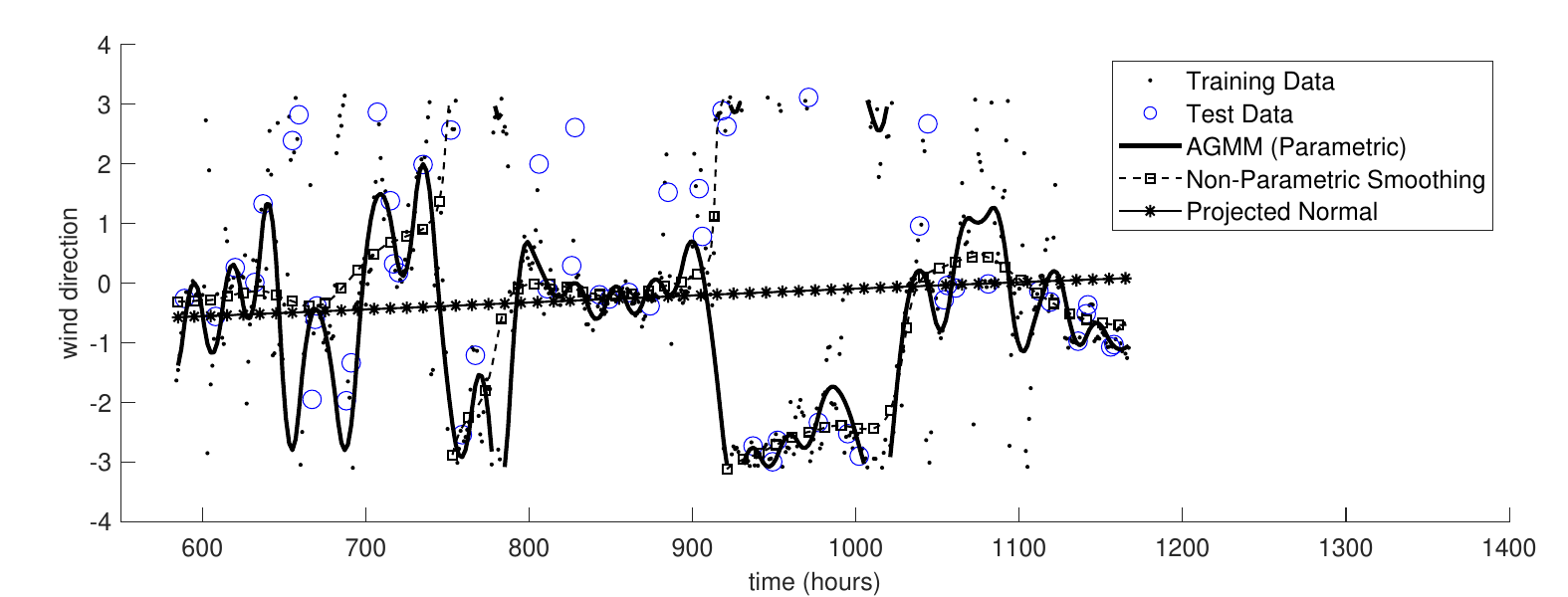}
	\includegraphics[width = \linewidth]{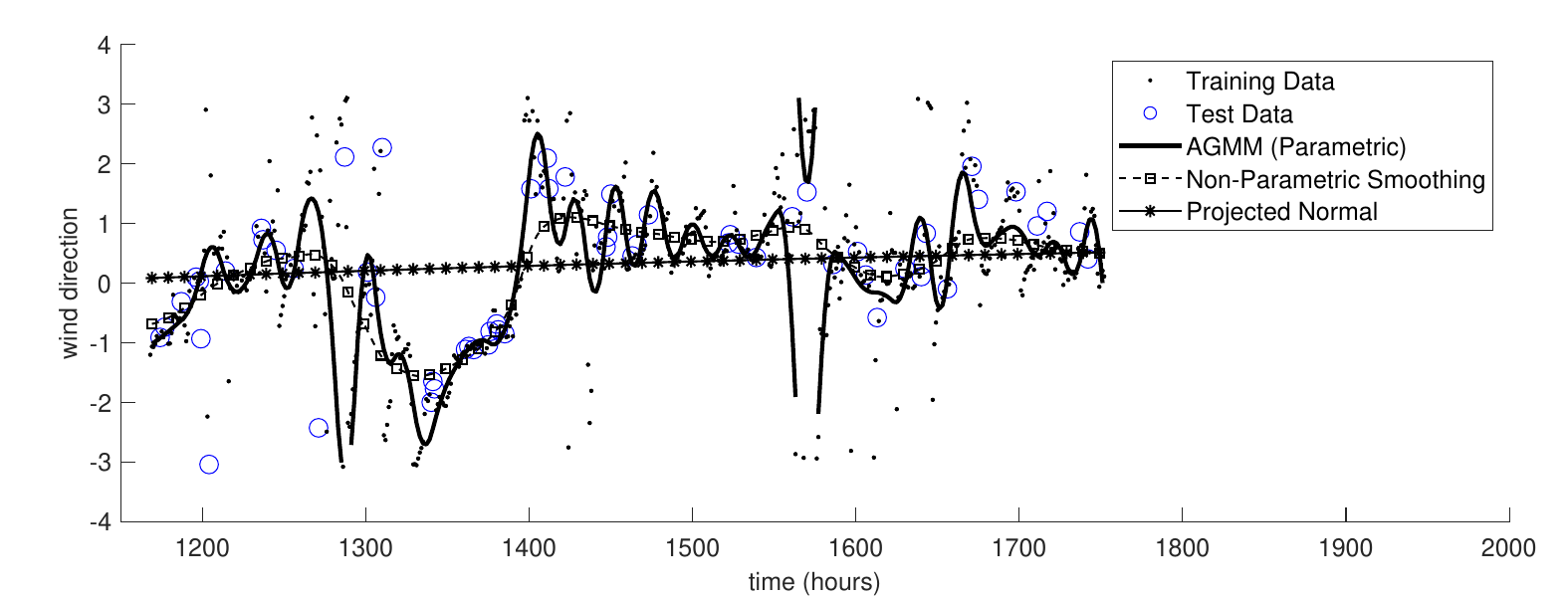}
	\centering
	\caption{Estimated mean functions of three compared methods for wind direction data.}
	\label{fig:realapp}
\end{figure}
We applied our proposed method for a problem of estimating wind directions at unobserved times given a number of observed wind directions. For the application, we used a dataset consisting of hourly wind directions measured at a weather station in Texas for from May 20 to July 31 2003. The dataset is a part of the Codiac data archive provided by the NCAR/EOL and is available at \url{https://data.eol.ucar.edu/dataset/85.034}. The dataset contains the 1,752 hourly average wind directions, among which 90\% was used for training data and the remaining 10\% was reserved for testing data. We applied the parametric AGMM with B-spline basis functions of degree three and evenly spaced knot locations, and the number of knots was determined to 150 by the 5-fold cross validation. The parameters of the AGMM were estimated using the Gibbs sampler with 30,000 iterations and 10,000 burn-in period.  

Figure \ref{fig:realapp} shows the estimated mean functions of the three compared methods with comparison to training data and test data; if it was plotted in a single plot, the plot extends very long along the x-axis, so we split the plot into three parts for better illustration. As shown in the training data, the wind directions fluctuates significantly over time with several significant directional jumps, which exhibits a functional pattern like a step function. The large variation mixed with circularity issues make it difficult to regress the data. For example, many observations of wind direction close to $\pi$ and -$\pi$ in time between 0 and 100 hours essentially show very uniform wind directions with some variations. However, due to circularity, the observations are split into the two ends of the $y$-axis, making complicated patterns. The proposed AGMM model captured such trend very accurately. The nonparametric smoothing approach produced the over-smoothed mean estimates, while the projected normal model did not capture the wind direction pattern. Table \ref{tbl:realdata} summarizes the numerical result. The overall mean circular error (MCE) of the AGMM estimation for the test data was 0.2204. The computation time was 2,572 seconds in a desktop computer with Intel Core i7-6600U CPU and 16 GB RAM. We also ran the non-parametric smoothing  \citep{di2013non} and the projected normal approach \cite{hernandez2017general} with the same training data. Their mean circular errors were not accurate as that of AGMM.

\begin{table}[h]
	\centering
	\begin{tabular}{|c|c|c|c|}
		\hline
		MCE & AGMM & Non-Parametric  & Projected Normal   \\ 
		(Time in seconds) & (Parametric) & Smoothing & Model \\ \hline
		Real Data &  0.2204  &  0.2576  &  0.5157 \\
		$n=1576$ &  2,572  &  3.4 &  579.4 \\
		         & 30,000 Iter & 5-Fold CV &  30,000 Iter \\ \hline	\end{tabular} 
	\caption {Total computations times. This table aims to provide a summary of computation time of all three methods on different examples. The projected normal model has considerably high computation time compared to the two other methods.}
	\label{tbl:realdata}
\end{table}

\section{Conclusion}
The AGMM model provides a novel modeling perspective to a circular response variable in a linear-circular regression problem. Many of the existing methods have regarded circular responses as projections of unobserved bivariate linear responses on the unit sphere, and the resulting projected normal distribution model was very expensive to estimate due to the modeling complexity. Compared to that, AGMM model is represented as a mixture of multiple linear models, which can be very effectively and efficiently estimated using many well established EM algorithms. The numerical performance of the AGMM model is also very promising. For five examples of different complexities, the new model outperformed the non-parametric smoothing and the projected linear model in terms of estimation accuracy. Its computation time was also very competitive. The computational competitiveness of the proposed approach allowed us to analyze a relative large dataset of wind direction measurements in time for the estimation of wind directions at unobserved times. The proposed approach completed the analysis in a reasonable time frame with much better estimation accuracy than the state-of-the-art.

\section*{Acknowledgment} 
We acknowledge support for this work from the AFOSR (FA9550-18-1-0144). 

\bigskip
\begin{center}
{\large\bf SUPPLEMENTARY MATERIAL}
\end{center}
\begin{description}
\item[MATLAB code for running numerical examples] MATLAB code containing code to perform the numerical examples in Sections \ref{sec:example} and \ref{sec:app}. (a zipped file)
\end{description}
\bibliography{bibfilev3}
\end{document}